Astro2020 Science White Paper

# Historical astronomical data: urgent need for preservation, digitization enabling scientific exploration

**Thematic Areas:** ☒Stars and Stellar Evolution, ☒ Planetary Systems

**Principal Author:**
**Name:** Alexei A. Pevtsov
**Institution:** National Solar Observatory, Boulder, CO, USA
**Email:** apevtsov@nso.edu
**Phone:** (303) 735-7728

**Co-authors: Elizabeth Griffin[2], Jonathan Grindlay[3], Stella Kafka[4], Jennifer Bartlett[5], Ilya Usoskin[6], Kalevi Mursula[6], Sarah Gibson[7], Valentín M. Pillet[1], Joan Burkepile[7], David Webb[8], Frédéric Clette[9], James Hesser[2], Peter Stetson[2], Andres Muñoz-Jaramillo[10], Frank Hill[1], Rick Bogart[11], Wayne Osborn[12], Dana Longcope[13]**

[1] National Solar Observatory, Boulder, CO, USA
[2] NRC Herzberg Astronomy and Astrophysics Research Centre, DAO, Victoria, B.C., Canada
[3] Harvard College Observatory, Cambridge, MA, USA
[4] American Association of Variable Star Observers, Cambridge, MA, USA
[5] U.S. Naval Observatory, Washington, DC, USA
[6] University of Oulu, Finland
[7] High Altitude Observatory, Boulder, CO, USA
[8] ISR, Boston College, MA, USA
[9] Royal Observatory of Belgium, Brussels, Belgium
[10] Southwest Research Institute, Boulder, CO, USA
[11] Stanford University, CA, USA
[12] Central Michigan University, Mt. Pleasant, MI USA, Member, AAS Working Group on Preservation of Astronomical Heritage, USA
[13] Montana State University, Bozeman, MT, USA

**Abstract:** Over the past decades and even centuries, the astronomical community has accumulated a significant heritage of recorded observations of a great many astronomical objects. Those records contain irreplaceable information about long-term evolutionary and non-evolutionary changes in our Universe, and their preservation and digitization is vital. Unfortunately, most of those data risk becoming degraded and thence totally lost. We hereby call upon the astronomical community and US funding agencies to recognize the gravity of the situation, and to commit to an international preservation and digitization efforts through comprehensive long-term planning supported by adequate resources, prioritizing where the expected scientific gains, vulnerability of the originals and availability of relevant infrastructure so dictates. The importance and urgency of this issue has been recognized recently by General Assembly XXX of the International Astronomical Union (IAU) in its Resolution B3: "on preservation, digitization and scientific exploration of historical astronomical data". We outline the rationale of this promotion, provide examples of new science through successful recovery efforts, and review the potential losses to science if nothing it done.

**The preservation of astronomy's historical data is critical for understanding our changing Universe, but those data are at increasing risk of becoming lost. Concerted efforts are needed urgently by the entire astronomical community to ensure that that does not happen. The data must be transformed digitally so that they can be accessible and used for the benefit of scientific exploration and thence of society.**

1. **The Problem**

Astrophysics – the archetypal observational science – is built upon pyramids of data collected over the years with a multitude of instruments, technologies and techniques from sites in many different areas of the globe. Assiduous observing has shown that everything actually changes at some level, manner or time-scale. Those changes contain clues to detailed intricacies of the formation and evolution of astrophysical bodies. Astronomy is well supplied with historical data: an estimated 4 to 7 million photographic observations of direct images and spectra, plus hand-drawings and written records. However, modern studies require data to be digital, whereas most of our historical data are analog only, and few observatories or individuals have the right tools or appropriate background knowledge to make correct use of such non-digital materials. Moreover, even early digital data are also at risk of being lost as their recording technology becomes obsolete. Without the necessary dedicated resources, knowledge about time-scales longer than two or three decades is effectively being denied us. This is extremely serious, as it curtails and biases the scope of astrophysics research.

Astronomical research would unquestionably benefit greatly from making the historical data available digitally. Having historical data in digital format would also enable amateurs and other citizen scientists to carry out very useful surveys and classifications, while the wealth of materials for teaching can open up new channels of instruction and hands-on experience for educators, all working towards the common goal, "The Universe: Yours to Discover". In the light of the unique scientific gains that could achieved by exploiting this threatened resource, a comprehensive and prioritized plan for conservation and digitization of our legacy data is urgently needed. The plan must be informed by (a) the potential scientific benefits of the data, (b) the risk associated with the deterioration of the records, and (c) the effort required to achieve full public access. Emerging computing and "big data" technologies are already leading to scientific breakthroughs by incorporating historical records.

Because analog data are cumbersome to access, their scientific potential is severely underutilized by researchers. But solution are not straightforward; translating photographic and paper records correctly into digital formats requires informed choices, specialized equipment, and human skills that are rapidly disappearing. Facilities and tools for preserving and digitizing the actual observations and other analog materials – in particular the associated logbooks and notebooks – are not available in most observatories. Add to this the fact that heritage archives, which belong to the community, are regarded as the responsibility of no-one, and we can understand how today's increasingly urgent situation has developed.

2. **What needs to be done**

Digitizing analog materials, in particular photographic plates, to extract the information correctly requires well-defined procedures but has to be guided by the anticipated scientific output. An inventory of heritage data is an essential preliminary requirement for assessing the size, condition and completeness of the various archives. Once the heritage data have been identified, it requires the development of a comprehensive and realistic plan for data preservation including digitization of data, creation of associated metadata, and the establishment of dedicated data centers to guarantee long-term access to the data and their discoverability. In the past, many important data sets were accumulated under PI-driven projects, but at termination the dataset could disappear, as nobody assumed responsible for the data curation. Access to historical data has certainly been hampered by their low level of discoverability. Without a long-term curation plan, even the digitized historical data may be lost owing to outdated data formats or ageing digital media (e.g., computer tapes, CDs). To be successful, the digitization and preservation of historical data requires close participation of all US funding agencies including NSF and NASA.

*A comprehensive long-term plan needs to be developed which establishes a clear path for historical datasets, prioritizes the data that need to be digitized and preserved, establishes the life-cycle of digitized data, identifies centers suitable for curating historical data, the data standards and the metadata. Most importantly, there is a strong need for dedicated funding for this activity, and support for pilot projects.*

### 3. New Science Enabled by Recent Projects

*3.1 DASCH*

Archives of the Digital Access to a Sky Century@Harvard (*DASCH*), the American Association of Variable Star Observers (AAVSO) and databases of solar observations extend back over 150 years and retain high scientific value. For instance, Arlot+ (2018) demonstrated that when modern astrometry is adopted for standards, the intrinsic accuracy of even the oldest images is actually significantly greater than previously thought. The ongoing digitization of the Harvard Astronomical Plate Collection is providing the world with a capability to study the sky systematically over 100-year time-scales (Grindlay et al. 2012). The team is

- Conducting the first long-term temporal variability survey over time-scales of days to decades,
- Investigating novae and dwarf novae distributions and populations in the Galaxy,
- Studying black-hole and neutron star X-ray binaries in outburst: constraining the BH, NS binary populations in the Galaxy,
- Investigating black-hole masses of bright quasars from long-term variability measures to constrain their characteristic shortest time-scales and thus their sizes,
- Determining black-hole properties in normally-quiescent galactic nuclei revealed by the optical flare resulting from the tidal disruption of a passing field star, and
- Finding unexpected classes of variables or temporal behavior of known objects, anticipating what PanSTARSS or LSST may see in greater detail but on shorter time-scales.

*3.2 AAVSO*

The AAVSO is building communities around historical and modern data. Since 1911 its international database has been collecting photometric data of more than 25,000 variable stars from the whole sky. It has also been digitizing early-20th century light-curves from the literature that were not digital, aspiring to provide a complete (as possible) history of brightness variations of those stars. The new information about stellar variability thereby revealed is leading to updated models of the underlying physical processes. Examples are legion; for example, Kiss+ (1999)'s detailed analysis of the long-term light curves of 93 red semi-regular variables revealed three distinct distributions in pulsating period space, representing long-term amplitude modulations, amplitude decreases and mode switching. Decade-long light-curves of dwarf novae reveal numerous new aspects of accretion physics. More than a century of SS Cyg data have revealed anomalous outbursts, super-outbursts and times of prolonged quiescence, triggering debates as to their nature and led to new models even before simultaneous X-ray observations suggested a two-component accretion flow model (e.g., McGowan+ 2003 and references therein). Long-term light-curves of individual objects reveal changes in their periodic behavior that fed missing links in stellar evolution. Results from a great many other studies of very long-term data (some spanning over 100 years) of novae and Mira variables are questioning our understanding of the evolution stages of novae, some of which may be the natural progenitors of SNeIa (e.g. Sabbadin+ 1983). Bodi+ (2016) combined *Kepler* data with long-term AAVSO light-curves to study pulsation stability in a number of RV Tauri stars, and discovered long and short-term pulsating cycles, which led to significant improvements in pulsation theories. A synergy between archival data and modern surveys is also demonstrated in the discovery of TYC-2505-672-1, which was shown to be a binary star system with a long period of 69.1 years, an eclipse duration of 3.45 years and an eclipse depth of 4 mag. (Rodriguez+ 2016; Lipunov+ 2016.)

*3.3 Heliophysics studies*

Solar activity waxes and wanes in 10- to 11-year cycles; this is now general public knowledge. However, we know it only because we could access archived long-term records. Thanks to those histories, we also know that the properties of solar cycles vary on timescales of hundreds of years and even millennia (Usoskin, 2017). Thus, some of the most important processes on the Sun may take decades if not centuries to reveal themselves (Owens, 2013). That long time-scale means that some issues are not resolved, or even identified, at the time when data are acquired.

Synoptic observations of solar activity, programs that span many years, feed future research to solve such issues. The project, entitled "Reconstructing Solar and Heliospheric Magnetic Field Evolution Over the Past Century" (by the International Team at the International Space Science Institute in Bern, Switzerland), uses (a) digitized, photographic, Ca K spectroheliograms from Kodaikanal Observatory (India, 1906 – 2000) and Mount Wilson Observatory (MWO, 1915 – 1985); and (b) sunspot field strengths (extracted from sunspot hand-drawings taken at MWO, 1917--present) to create pseudo-magnetograms of the solar surface from 1906 until the early 1970s (beginning of direct systematic magnetogram observations; Pevtsov+ 2016). These pseudo-magnetograms are employed as input to a modern flux transport model to reconstruct the evolution of weak magnetic fields outside active regions and the polar magnetic field (Virtanen+ 2017). Having "pseudo magnetograms" of the solar surface would enable one to reconstruct and model several key parameters of space weather such as the total solar irradiance, the solar wind and the heliospheric magnetic field over the last 100 years -- well before space exploration began. The data will enable the use of modern physics-based models for interpreting long-term changes in solar activity, and their effects upon solar-planetary interactions. In another recent example, Svalgaard (2016) used historical records of the geomagnetic field from different stations (46 million hourly measurements) to reconstruct the solar extreme UV flux from 1740 –2015.

Understanding solar activity will benefit enormously from studies of historical records of solar spectra, magnetograms, faculae and variations in sunspot counts.

*3.4 Other Projects*

*(A) Photometry.* Large surveys on small telescopes have proved hugely fruitful. The Sonneberg Plate Patrol, for example, which operated from 1927 to the late 1980s and has a database of some 275000 plates, led to the discovery of about 25% of the then known variables stars. Digitizing such material to the spatial resolution of the emulsion grain size (as done by *DASCH*) could find more, particularly if they also include full photometry (e.g. using Sextractor) and astrometry (vs. GaiaDR2) as is done for *DASCH*. Collections like those will also fill in gaps in other important series (such as *DASCH*, which is marred by the 'Menzel Gap' from ~1955--70; Johnson 2007). Particularly valuable would be digitized data (and reductions) from southern hemisphere collections to complement northern plate-series.

*(B) Astrometry.* It was confidently predicted that the precision of Gaia would prove to be so high that historical sets of plates like the *Carte du Ciel* would be unnecessary. However, it has now been shown (Arlot+ 2018) that the published precision of the historical plates was governed by poor positions for the references stars used, and that when Gaia positions for the same stars are used then the precision of the measured images on the plates increases by a factor of ~10 (to 30 – 60 mas), thereby increasing very significantly the re-use value of those plates for establishing long-periodicities, particularly in studies of solar-system objects.

*(C) Spectroscopy.* Cataclysmic Variables, flare stars, chromospherically-active stars, novae, supernovae, spectroscopic and eclipsing binaries and chromospheric-eclipsing systems have been the traditional hunting-ground of amateurs, whose efforts have been collected systematically for research since 1911 through the AAVSO. However, there has been little routine follow-up of those objects spectroscopically to support the photometry per se, though numerous individual studies of those variables have been carried out with spectroscopy in a variety of modes, from objective-prism or Cassegrain spectroscopy (for broad classification and for estimating spectral energy distributions) to medium- and high-dispersion coudé grating spectroscopy (for accurate stellar abundances, tomography and planetary studies), thus servicing all kinds of research. Binaries having periods longer than about 20 years can barely be analyzed adequately from modern archived data, while period modulations that disclose the presence of extra bodies in the systems need spectroscopy extending over multiple periods. Spectra of the illustrious 'chromospheric eclipses', manifested by select eclipsing stars like ζ Aurigae, show that the chromosphere of a late-type giant is highly dynamic and variable. The A-type stars are yielding valuable measurements of stellar rotation when surface spots modify the profiles of absorption lines.

All of these studies, any many similar ones, are scarcely possible without the rich databases of spectra that encompass adequately long time-spans that only historical data can offer. The results are immediately benefiting stellar physics, and in particular stellar-evolution theory.

*3.5 Trans-Disciplinary Research*

Instances in which astronomical data can yield direct new knowledge for other sciences are spectacular, though not yet plentiful. One fascinating re-use of historical astronomical spectra was the attempt to monitor concentrations of the Earth's stratospheric ozone by analyzing historical ultra-violet spectra of hot stars. Atmospheric ozone has been monitored (in Switzerland) since 1926, but there were no suitable back-up measurements until the 1950s. Attempts to study – and more importantly, to repair -- the 'ozone hole' discovered in the 1970s (Farman+ 1985) depend on information gleaned from the relatively noisy data of the early recordings. Other sources are therefore needed, and astronomy offers a unique advantage because every ground-based observation inevitably bears the signature of the Earth's atmosphere too. Digitizing and analyzing suitable historical spectra can provide precious information, as demonstrated in a pilot study (Griffin, 2005, 2006). Making more spectra available digitally will enable a far broader attack to be made on this vital problem and perhaps others yet to be formulated.

*3.6 Supportive Proxy Data*

Historical observations and/or records of aurora borealis/australis, supernovae, planetary transits, etc., have been used through past centuries as indirect proxies for astronomical phenomena. Those data are spread over non-astronomical manuscripts and chronicles, and are sometimes either totally unknown to the astronomical community or can only be found in an erroneous or incomplete interpretation, which has been subsequently reproduced in modern literature with a lost reference to the original source. Such data need to be systematized, verified and catalogued with an access to digitized copies of the original sources. Such projects could inspire new collaboration with historians, geographers and other disciplines.

*3.7 The Broader Impact of Historical Data*

The preservation and scientific exploration of historical data already has a broad impact well beyond the immediate astronomical community. Its demands will continue to generate unique experiences in stellar variability, time-domain science, software skills, Big Data analytics, data management, hardware design and technology, and publicizing science to non-astronomers. It also offers special value as an educational tool. Associated projects will enable new collaborations between astronomers, historians, archivists, librarians, students, and amateurs engaged in public outreach and citizen science.

**4. The Need for Community Action**

Time is not on our side. Plates age, though usually fairly slowly, but handwritten records on paper (including metadata in logbooks) and hand-drawn observations are ageing rapidly and soon many will no longer be legible. All are vulnerable to damage or destruction by natural hazards (floods, fire, earthquakes), by vermin, and – worse – through human ignorance; there are already sad cases when a lack of recent use has been interpreted too hastily as a lack of interest, and a valuable collection has been discarded. The earliest observations recorded digitally are also now at risk as the technology needed to read their media are discarded as obsolete. The extant wisdom in the community will also be lost before long, as people retire or disappear from the scene; losing it entirely will prove disastrous in the case of many data sets. The time to act is therefore *now*. The astronomical community must unite to save its inherited data – or face losing them all. In fact, pockets of concern to save data are already beginning to appear, though not in any coordinated manner. As a community we need to consolidate those beginnings.

The importance and urgency of the preservation and scientific exploration of historical data was recognized in 2018 by the astronomical community by the passing of IAU Resolution B3: "on preservation, digitization and scientific exploration of historical astronomical data". That Resolution is binding upon the entire Union, which is thereby agreeing to assist in urging its realization. We are therefore calling all astronomers – whether directly or only indirectly involved in variability studies or in acquiring data for them – to grasp this opportunity to inform others, to link to other groups, to broadcast the situation to outsiders (the press, historians, other professionals, science educators) and to support every effort to develop a roadmap for preservation of historical data and raise the funding which will be needed to carry out local or more distant preservation or digitization efforts. Astronomers must speak with one voice in this issue. It affects us all, it is the responsibility of all, and it will ultimate be to the benefit of all.